\documentclass[aps,showkeys,showpacs,superscriptaddress,twocolumn,floatfix]{revtex4}
\usepackage{graphicx}
\usepackage{amsmath}
\usepackage{amsfonts}
\usepackage{color}

\begin{document}

\title{The risk of extinction --- the mutational meltdown or the overpopulation}

\author{Krzysztof Malarz}
\homepage{http://home.agh.edu.pl/malarz/}
\email{malarz@agh.edu.pl}

\affiliation{
Faculty of Physics and Applied Computer Science,
AGH University of Science and Technology,\\
al. Mickiewicza 30, PL-30059 Krak\'ow, Euroland
}

\affiliation{
Institute of Theoretical Physics,
University of Cologne, D-50923 K\"oln, Euroland
}

\date{\today}

\begin{abstract}
The phase diagrams survival-extinction for the Penna model with parameters: (mutations rate)-(birth rate), (mutation rate)-(harmful mutations threshold), (harmful mutation threshold)-(minimal reproduction age) are presented. 
The extinction phase may be caused by either mutational meltdown or overpopulation.
When the Verhulst factor is responsible for removing only newly born babies and does not act on adults the overpopulation is avoided and only genetic factors may lead to species extinction.
\end{abstract}

\pacs{
87.23.Cc, %% Population dynamics (ecology)
05.45.-a  %% Catastrophe theory
}

\keywords{Monte Carlo simulation, phase transition, ageing, catastrophe}

\maketitle

%% ############################################################################
\section{Introduction}
%% ############################################################################

A species becomes {\bf extinct} when the last existing member of that species dies.
Extinction therefore becomes an extreme event after which no surviving specimens are able to reproduce and create a new generation.
There have been at least five mass extinctions in the history of life in which many species have disappeared in a relatively short period of geological time.

The classical ``Big Five'' mass extinctions identified by Raup and Sepkoski \cite{raup} were approximately 444 ({\em End Ordovician}), 367 ({\em Late Devonian}),  255 ({\em End Permian}), 200 ({\em End Triassic}) and 65 ({\em End Cretaceous}) million years ago.
These mass extinctions may have been caused by impact events, climate changes, volcanism, $\gamma$-ray burst and plate tectonics or the mixtures all of them.
During {\em End Permian} extinctions --- the worst in the Earth history --- 96\% of marine species and 70\% of land ones (including plants, insects and vertebrates animals) were killed. 
The most recent, the most famous and most rapid mass extinction (65 million years ago, at the end of the Cretaceous period) wiped out all non-avian dinosaurs and it was caused probably by a cosmic collision with an asteroid or comet of ten kilometers in diameter \cite{wikipedia}.

We are currently in the early stages of a human-caused mass extinction, known as the {\em Holocene extinction} event.
Probably, up to twenty percent of all living species could become extinct within thirty years (by 2028).
Wilson estimated \cite{wilson} that if current rates of human destruction of the biosphere continue, half of all species of life on Earth will be extinct in hundred years \cite{wikipedia}.

On the other hand even without external tragedies the fate of populations may be quite unpleasant when an organism's number exceeds the carrying capacity of its ecological niche.
Such situation is termed as {\bf overpopulation}.
Thomas Malthus argued that if left unrestricted, human populations would continue to grow until they would become too large to be supported by the food grown on available agricultural land \cite{malthus}.
He proposed that, while resources tend to grow linearly, population grows exponentially.
At that point, the population would be restrained through mass famine and starvation.
Malthus argued for population control, through ``moral restraint'', to avoid this happening.
As the population of a species exceeds the amount of available resources, it decreases, sometimes sharply, since the lack of resources causes mortality to increase \cite{wikipedia}.

Could a population become extinct due to genetic factors?
The answer is probably positive.
The Darwinian evolution theory \cite{darwin} predicts genetic modification of individuals genome via mutation process and slow change one species to another.
This change (called {\bf speciation} \cite{speciation,eigen-MM,eigen}) and the newly created species will be accepted by Nature through the natural selection procedure.
In this way mutations --- usually regarded as harmful --- may become crucial for new species creation and evolution.
However, sometimes too many genetic modifications yield genetic death of the individual. 
Sometimes it happens for all individuals in a given population ---  we call that the {\bf mutational meltdown} \cite{mmelt}.

%% ############################################################################
\section{The Penna model}
%% ############################################################################

The Penna model of biological ageing \cite{penna} is devoted to reproduce single-species population dynamics for genetically heterogeneous individuals represented only by their genotype --- the $N_{bit}$ long binary computer word.
The population fate is assumed to be governed by accumulation theory \cite{review}, which claims that random hereditary deleterious mutations accumulate over several generations in the species genomes.
These bad mutations are represented as ``1'' in the genome.
The time is measured by discrete variable $t$.
In each time step (which may be treated as a ``season'' and may have different meanings for different species), for each individual of age $a$, the number of bits set to one in the first $a$ positions in the genome is calculated. 
These ``1'' in position $1\le i\le a$ are treated as active mutations.
If the number of active mutations is greater or equal to the threshold value $T$ an individual dies.

Individuals compete among themselves for resources (for example food and territory): each of them may be removed from the population with probability $N(t)/N_{max}$, where $N_{max}$ represents the maximal environmental capacity and $N$ is the current number of individuals. 
This process is independent on individual age and health.
The term $N(t)/N_{max}$ is termed a Verhulst's factor \cite{verhulst}.
The Verhulst factor in natural way limits exponential explosion of the population size to the infinity, to avoid the Malthus catastrophe.

The random removing of organisms due to their competing for resources is not very well justified biologically: we feel that the well-fitted individuals should be more resistant for random removing --- which mimics competition --- that ill-fitted ones.
Thus, to keep the population finite and avoid accidentally killing the best-fitted organisms, the Verhulst factor may act only on newly born babies and not on adult ones \cite{martins00}.
Here, however, the idea of massacre of the innocents --- proposed by Herod the Great \cite{herod} --- is applied to one season old children and not to children who are two years old and under.

The population reproduces asexually employing cloning mechanism.
If the individual is older than a minimum reproduction age $R$ but before loosing its fertility --- which happens at age $E$ --- it clones itself with probability $b$ producing $B$ offspring.
The cloning process is not perfect: during replication parent's genome is exposed to harmful mutations (``0''$\to$``1'') which occur with probability $m$ at $M$ randomly selected positions in genome.
The length of genome $N_{bit}$ restricts maximal age of individuals.

The extinction process within the frame of the Penna model was discussed rather seldom. 
In the literature known to us only the papers by Maksymowicz {\em et al} \cite{maksymowicz99}, Bernardes \cite{bernardes95}, P\'al \cite{pal96} and Fehsenfeld {\em et al} \cite{fehsenfeld03} deal with species extinction.
Some possible suggestions about mutational meltdown are also available in Refs. \cite{malarz00,malarz04,laszkiewicz}.
On the other hand the rapid survival rate decreases for semelparous organisms (like Pacific salmon \cite{salmon}) or the influence of an overfishing on northern cod fish population \cite{cod} were one of the first applications of the Penna model \cite{oldbook}.

\bigskip

In this paper we would like to draw a phase diagrams in the space of the Penna model \cite{penna} control parameters which for initially large populations separate survival from extinction.

%% ############################################################################
\section{The results}
%% ############################################################################

We start our simulation with $N_0=3\cdot 10^5$ individuals with perfect genomes ($N_{bit}$ zeros) and with environmental capacity $N_{max}=10^6$.
We allow individuals to reproduce until their death ($E=N_{bit}=32$).

\begin{figure}
\includegraphics[width=0.49\textwidth]{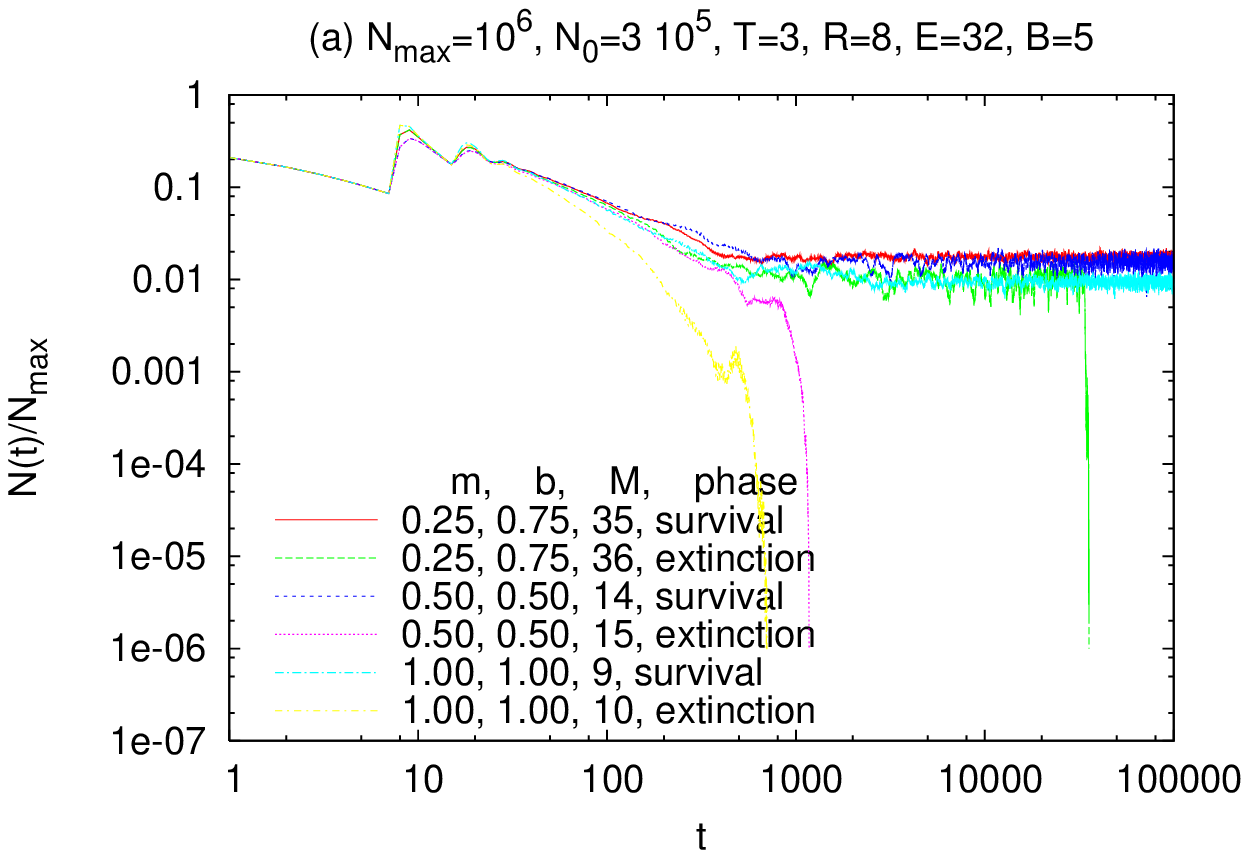}\\
\includegraphics[width=0.49\textwidth]{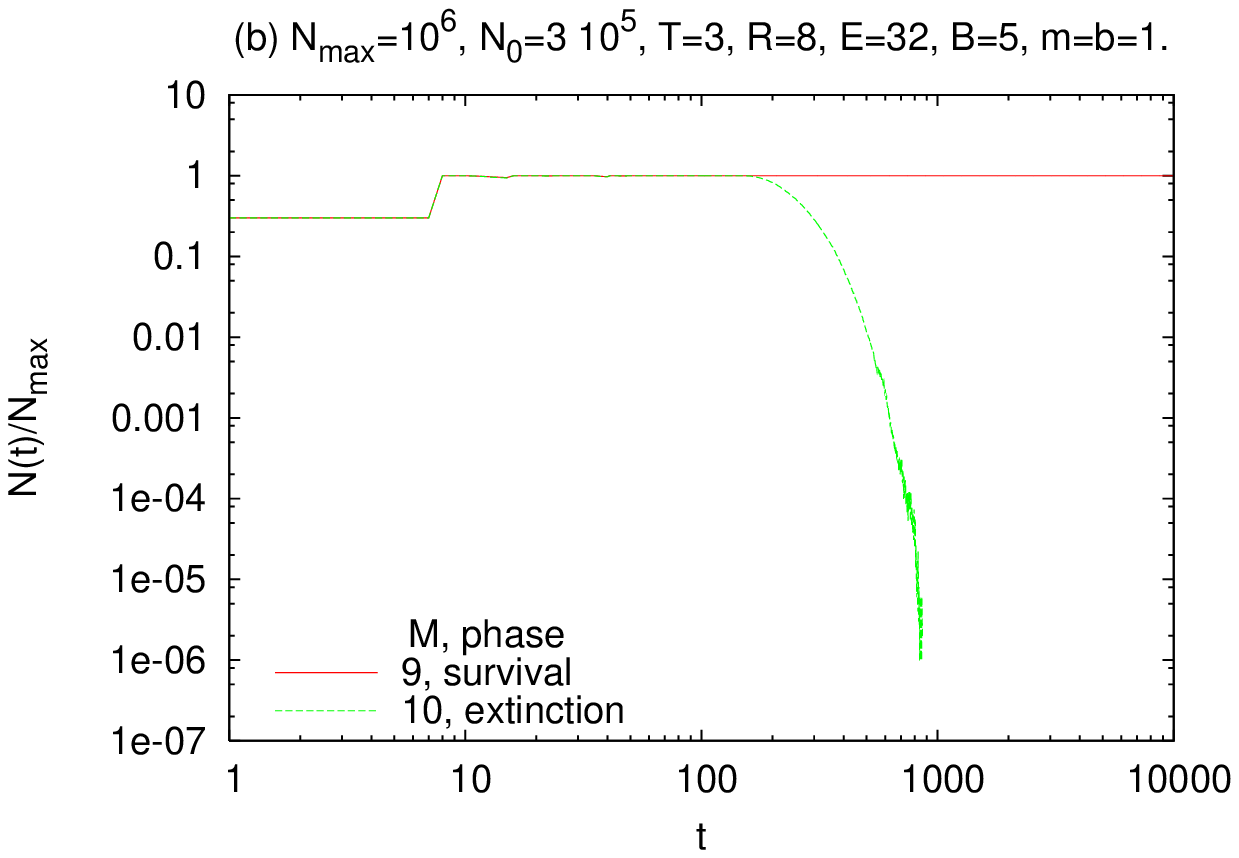}
\caption{A few examples of population size dynamics for different mutation and birth rates ($M$, $B$) and probabilities ($m$, $b$).
The subsequent pairs of curves have identical set of parameters except of mutation rates which are $M$ and $M+1$.
The Verhulst's factor applies either to (a) all individuals or (b) only to newly born babies.}
\label{N_vs_time}
\end{figure}

\begin{figure}
\includegraphics[width=0.49\textwidth]{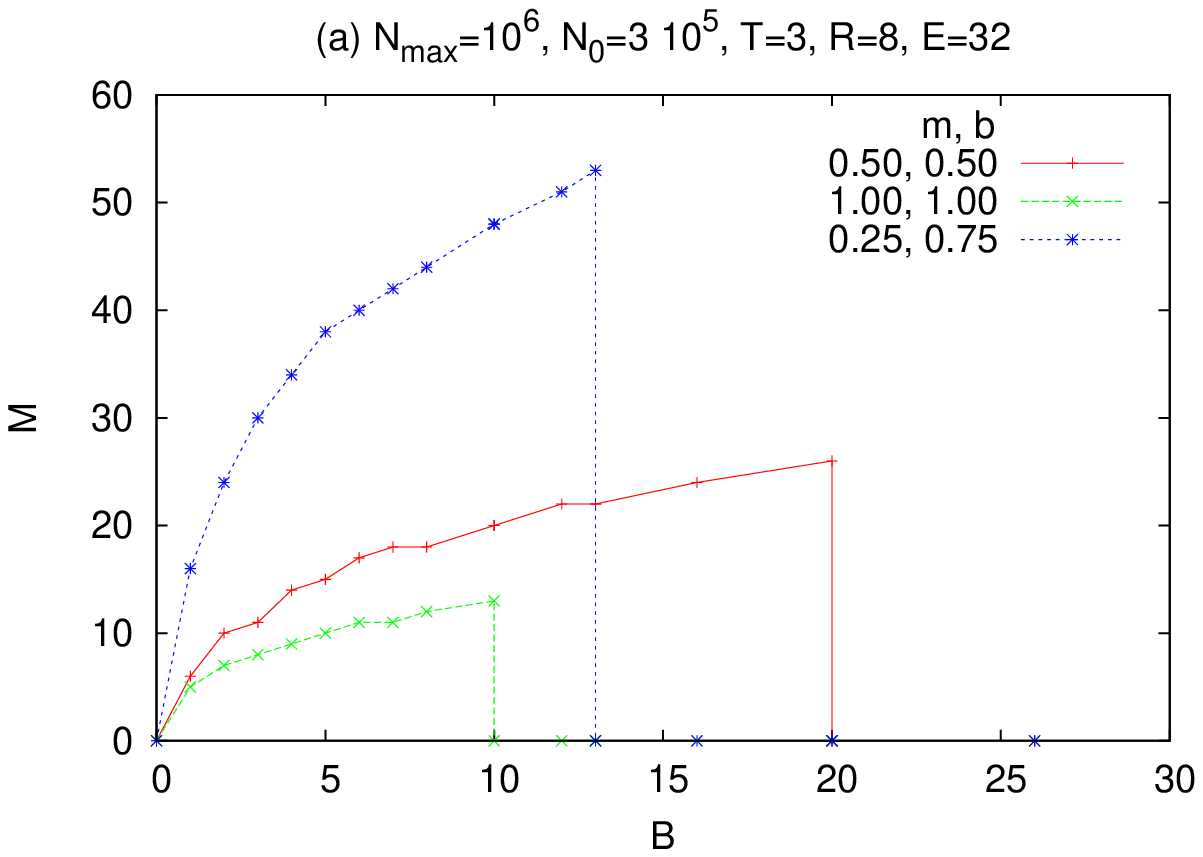}\\
\includegraphics[width=0.49\textwidth]{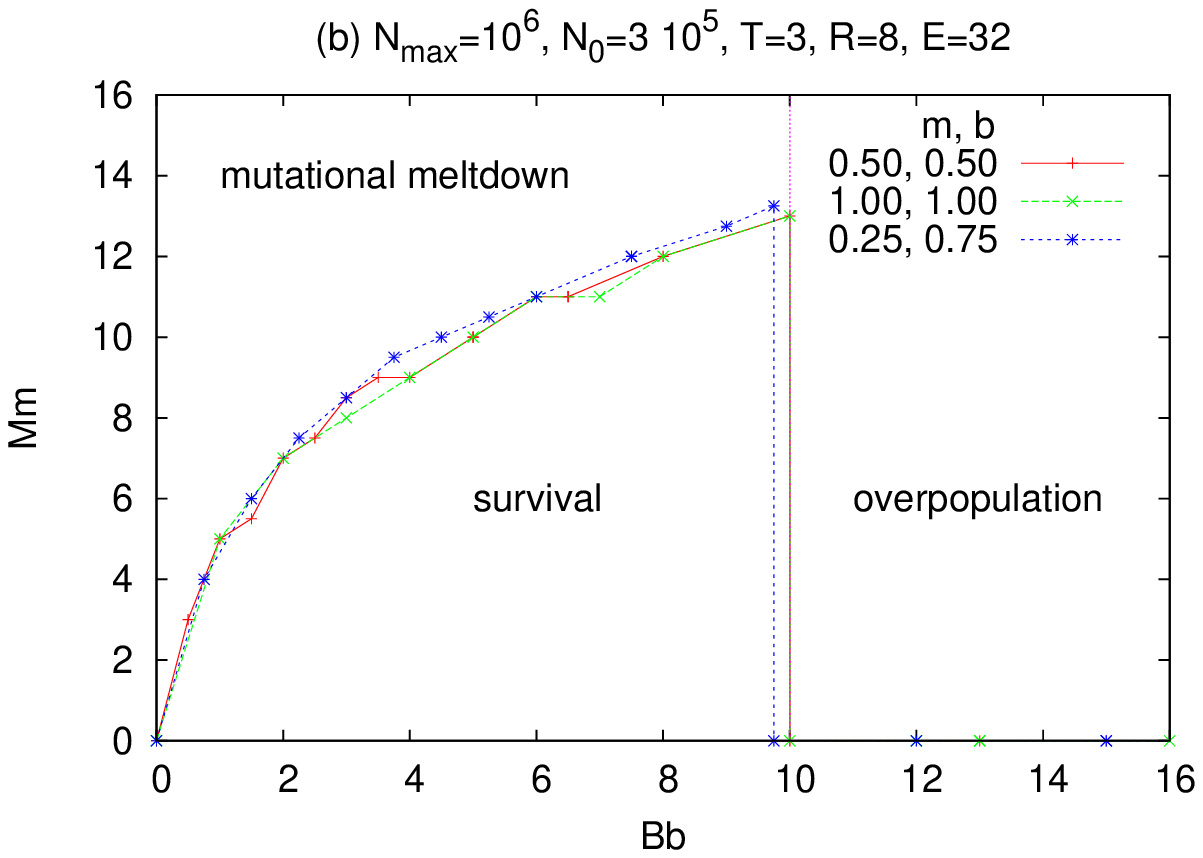}\\
\includegraphics[width=0.49\textwidth]{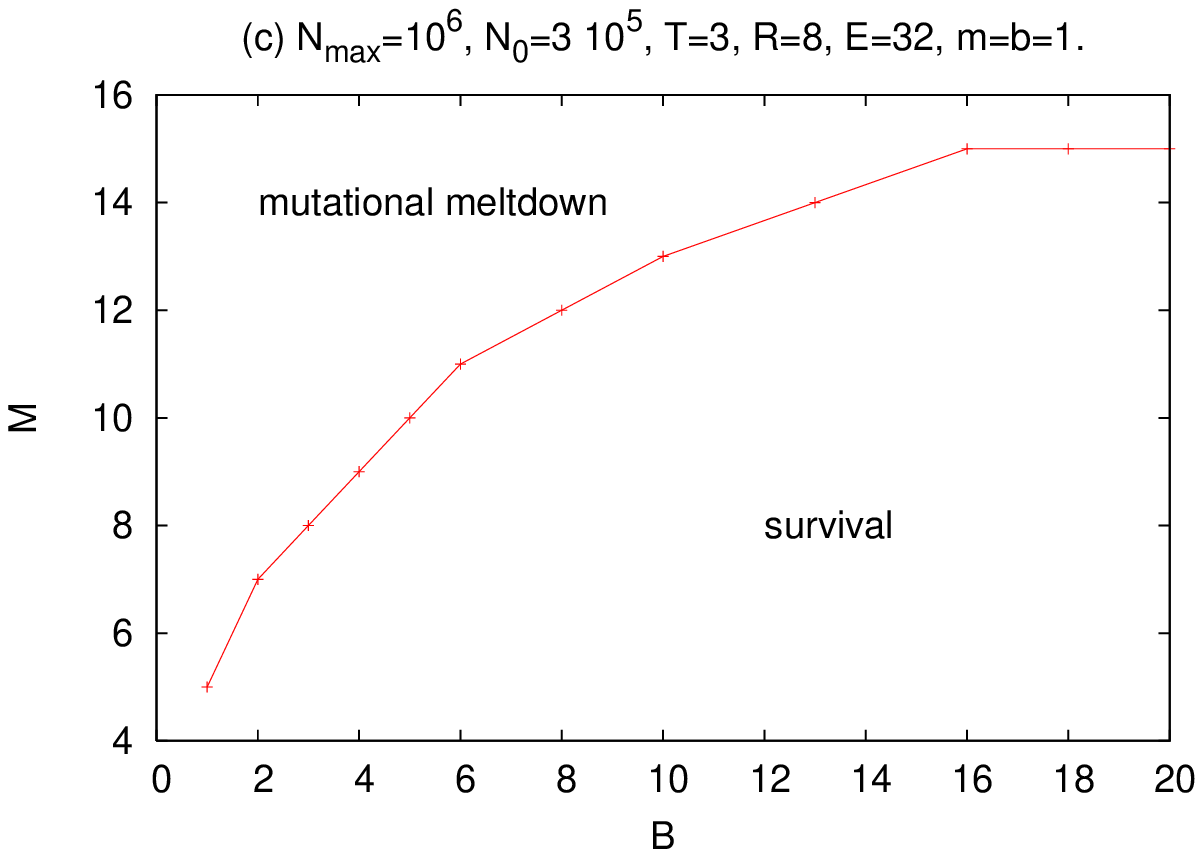}
\caption{The phase diagram survival-extinction for (a) various values of the mutation and birth rates ($M$, $B$) and probabilities ($B$, $b$) 
and (b) for average numbers of mutations $Mm$ and births $Bb$ per ``season'' and per individual.
Below arc-like lines the species survival is possible.
At and above them individuals die  due to too many mutations pumped into their genome.
On the right side of the vertical line the extinction process is caused by overpopulation.
(c) The latter may be avoided when Verhulst's factor acts only on newly born babies.}
\label{MB}
\end{figure}

\begin{figure}
\includegraphics[width=0.49\textwidth]{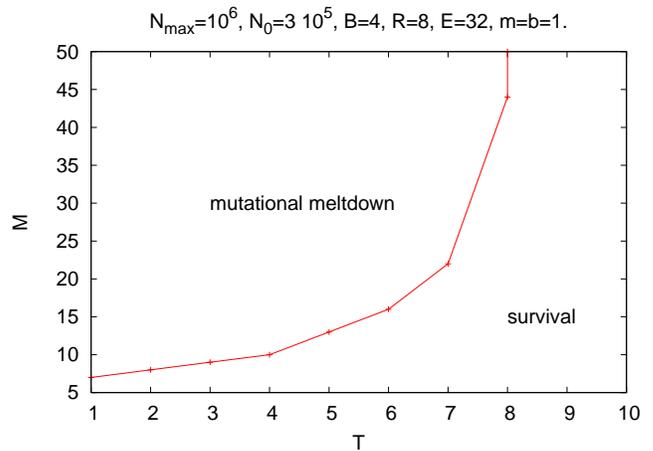}
\caption{The phase diagram survival-extinction on (harmful mutation threshold $T$)-(mutation rate $M$) plane.}
\label{TM}
\end{figure}

\begin{figure}
\includegraphics[width=0.49\textwidth]{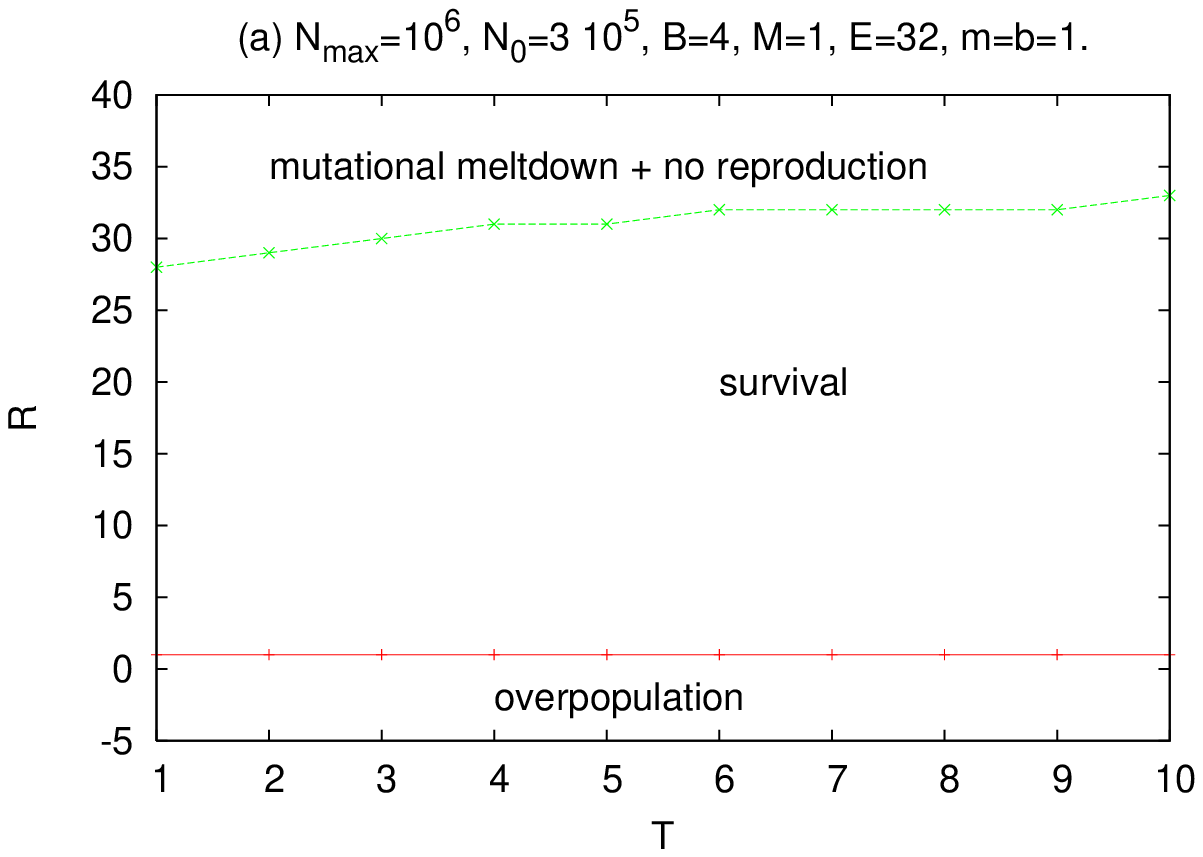}\\
\includegraphics[width=0.49\textwidth]{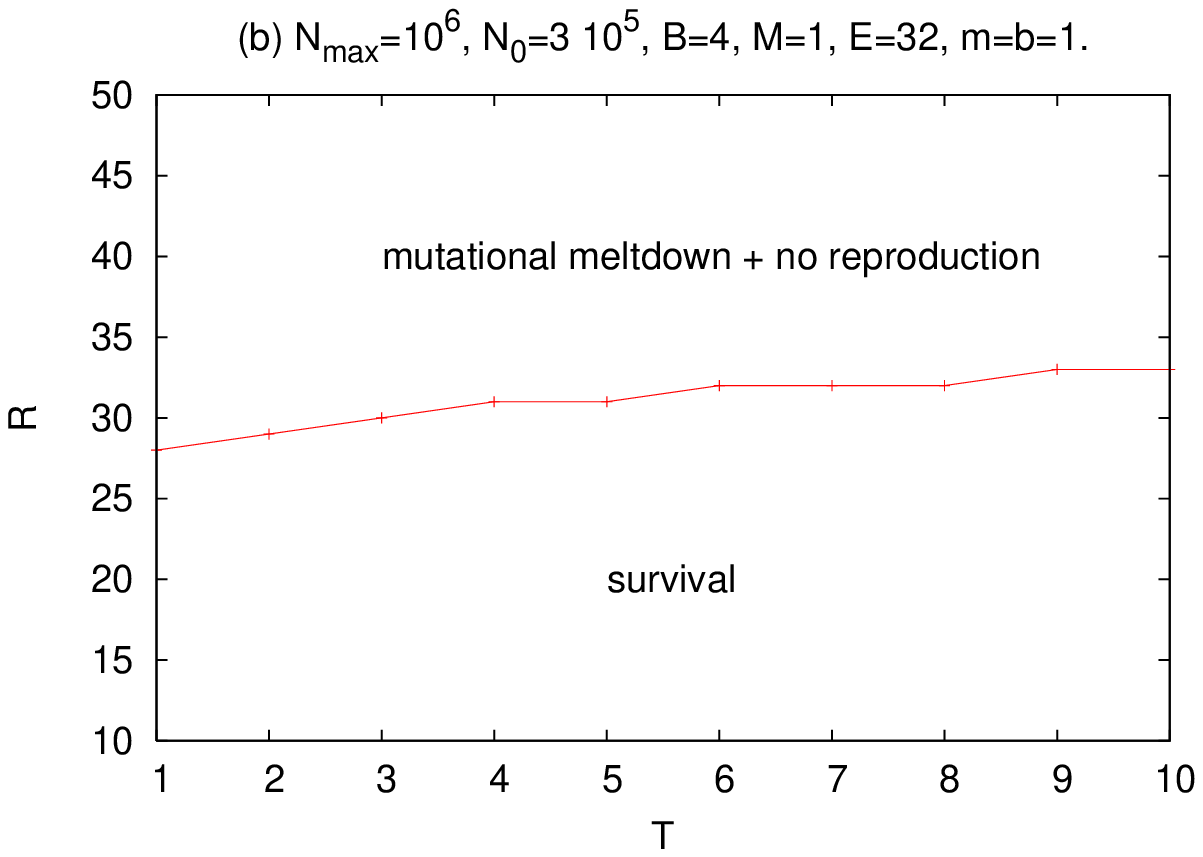}
\caption{The phase diagram survival-extinction on (harmful mutation threshold $T$)-(minimal reproduction age $R$) plane for (a) Verhulst's factor acting on all individuals and (b) only on newly born babies.
In the latter case the overpopulation is avoided.}
\label{TR}
\end{figure}

In Fig. \ref{N_vs_time} examples of the population density $N(t)/N_{max}$ are presented.
Subsequent pairs of curves correspond to sets of $(m,b)$ parameters (0.25, 0.25), (0.5, 0.5) and (1, 1) from top to bottom, respectively.
The rest of parameters is fixed except the mutation rate $M$.
For fixed set of parameters ($T$, $R$, $E$, $B$, $m$, $b$) the time evolution of population density differs qualitatively for mutation rate $M$ (survival) and $M+1$ (extinction).
Such and similar differences allow to construct phase diagrams presented in Figs. \ref{MB}, \ref{TM} and \ref{TR}.
In Fig. \ref{N_vs_time}(b) the Verhulst factor reduces only a birth probability $b \to b[1-N(t)/N_{max}]$ and not the total number of individuals (see Fig. \ref{N_vs_time}(a)).

%% ----------------------------------------------------------------------------
%% ----------------------------------------------------------------------------

%% ----------------------------------------------------------------------------
%% ----------------------------------------------------------------------------

In Fig. \ref{MB} the phase diagrams survival-extinction for various values mutation and birth rates ($M$, $B$) and probabilities ($B$, $b$) and for average numbers of mutations $Mm$ and births $Bb$ per ``season'' and per individual are presented.
The phase transition border for average values of number of mutations $Mm$ and average number of offsprings $Bb$ shows some kind of scaling universality.
Obviously, the larger mutational rate is the higher is the birth rate necessary to keep the population alive.
On the other hand, the population size $N(t)$ is restricted to the environmental capacity $N_{max}$ of the niche where this population lives: too large fertility of individuals may easily lead to species extinction due to overpopulation \cite{bernardes98}. 
The overpopulation effect may be avoided when Verhulst factor acts only on the newly born babies \cite{martins00} (see Fig. \ref{MB}(c)).

%% ----------------------------------------------------------------------------
%% ----------------------------------------------------------------------------

In Fig. \ref{TM} the phase diagram survival-extinction for various values mutation rates $M$ and the critical mutations threshold $T$ is presented.
The mutational pressure shifts damaged genes (positions of ``1'' in genome) in that way that only $T-1$ mutations were located in the first $R$ positions of genome.
This ensures chance of reproduction for individuals \cite{review}.
The population is stable for $T>R$ independent of mutation rate $M$ (see Ref. \cite{bernardes98}).
For a lower threshold $T<R$ the critical mutation rate $M$ exists above which the population dies out due to mutational meltdown.

%% ----------------------------------------------------------------------------
%% ----------------------------------------------------------------------------

The phase diagram on the plane of minimal reproduction age $R$ versus harmful mutation threshold $T$ seems to show trivial behavior: when you allow individuals to reproduce immediately after they were born --- the number of individuals explode exponentially.
When the population size crosses the border of environmental capacity $N_{max}$ the Verhulst's factor will remove all individuals due to overpopulation.
On the other hand when you shift the minimal reproduction age $R$ almost to infinity (which here means $N_{bit}$) the extinction will occur (see Fig. \ref{TR}) as individuals will have no chance for reproduction.
And again: the overpopulation effect may be avoided when one applies the Verhulst factor only on the infants.

It seems that the transitions remain unchanged when the simulation time or the population size are enlarged.

%% ############################################################################
\section{The Conclusions}
%% ############################################################################

The subject of extinction is still attractive for scientists (see Refs. \cite{mmelt,extinction,overpopulation} for the most recent papers in the field.)

The influence of genetic factors on population extinction was numerically investigated also earlier \cite{eigen-MM} for the Eigen model \cite{eigen} of evolution, where the population does not possess age structure.
The Penna model with its set of control parameters is very flexible and thus almost any of experimentally observed effect for real population may be successfully reproduced \cite{review}.

Some trivial assumptions of the values of the Penna model parameters may lead to species extinction.
For example, when one assumes individuals infertility ($B=b=0$) or one forgets to give individuals a chance for reproduction ($E<1$ or $R>N_{bit}$) the initial population, even when huge, will die out.

In this paper we show how the populations may be fragile when the main factor responsible for removing individuals is either too rapid growth of the population (which results in exhausting resources necessary for life) or genetic mutational meltdown of the population.
In both cases the phase transition extinction-survival is observed.
When the environmental capacity is finite --- and the Verhulst's factor guards maximal environment capacity border --- even without any errors in reproduction (no mutations) the population may explode in one ``season'' much over allowed the limit and die out due to starvation.
The effect of overpopulation may be removed when Verhulst factor kills only babies.
On the other hand, for fixed average reproduction rate $Bb$ the critical mutation rate $Mm$ may be found above which population cannot have stable and non-zero size. 

The latter means, that even without catastrophe in the cosmic scale the last examples of the endangered species --- for example
the giant panda ({\em Ailuropoda melanoleuca}),
the red panda ({\em Ailurus fulgens}, better know as Firefox),
Hickman's potentilla ({\em Potentilla hickmanii}),
the San Francisco garter snake ({\em Thamnophis sirtalis tetrataenia})
\cite{redlist} --- may be finally removed from the Earth and meet the fate of phoenix, dragons, gargoyles and trolls.
But simulated extinction of such small populations was already investigated in details earlier \cite{pal96}. 

Last but not least: extinction not always must mean bad thing as the extinction of one species usually open the free way for evolution of another one \cite{speciation,review}.
The most famous mass extinction --- which killed all dinosaurs 65 million years ago --- perhaps was not very polite for these huge lizards (particularly from their point of view), but prepared place for a race to which author and anonymous referee(s) of this paper belong to.

%% ############################################################################
\begin{acknowledgments}
%% ############################################################################
Author is grateful to Dietrich Stauffer for his hospitality in K\"oln and to Volker Jentsch of the Complexity Centre at Bonn University for suggesting this work.
\end{acknowledgments}

%% ############################################################################

\end{document}